\documentclass[conference]{IEEEtran}
\IEEEoverridecommandlockouts
\usepackage{amsmath,amssymb,amsfonts,graphicx,mathtools,nccmath,amsthm,float}
\usepackage{algorithm,multicol,multirow,tikz,cite,array,booktabs,url}
\usepackage{caption}
\usepackage{algpseudocode,algorithmicx}
\usepackage{cite}
\usepackage{textcomp}
\usepackage{xcolor}
\usepackage[inline]{enumitem}
\usepackage{subcaption}
\usepackage[cal=cm,scr=euler]{mathalpha}

\graphicspath{{figures/}{./}}
\DeclareGraphicsExtensions{.pdf}

\def\BibTeX{{\rm B\kern-.05em{\sc i\kern-.025em b}\kern-.08em
    T\kern-.1667em\lower.7ex\hbox{E}\kern-.125emX}}

\newcommand{\vect}[1]{\boldsymbol{\mathbf{#1}}}

\DeclarePairedDelimiter{\norm}{\lVert}{\rVert}
\DeclarePairedDelimiter{\abs}{\lvert}{\rvert}

\algnewcommand{\LineComment}[1]{\State \(\#\) #1}
\algnewcommand\algorithmicinput{\textbf{Set}}
\algnewcommand\Set{\item[\algorithmicinput]}
\algnewcommand\algorithmicinitial{\textbf{Initialize}}
\algnewcommand\Initialize{\item[\algorithmicinitial]}

\let\oldReturn\Return
\renewcommand{\Return}{\State\oldReturn}

\begin{document}

\title{Robust Consensus-Based Distributed Beamforming for Wideband Cell-free Multi-RIS MISO Systems}
\author{\IEEEauthorblockN{Konstantinos D. Katsanos$^1$ and George C. Alexandropoulos$^{1,2}$}
\IEEEauthorblockA{$^1$Department of Informatics and Telecommunications,
National and Kapodistrian University of Athens\\
$^2$Department of Electrical and Computer Engineering, University of Illinois Chicago, USA\\
emails: \{kkatsan, alexandg\}@di.uoa.gr\vspace{-0.5cm}
\thanks{This work has been supported by the Smart Networks and Services Joint Undertaking (SNS JU) projects TERRAMETA and 6G-DISAC under the European Union’s Horizon Europe research and innovation programme under Grant Agreement No 101097101 and No 101139130, respectively.}
}}

\maketitle

\begin{abstract}
The cell-free networking paradigm constitutes a revolutionary architecture for future generations of wireless networks, which has been recently considered in synergy with Reconfigurable Intelligent Surfaces (RISs), a promising physical-layer technology for signal propagation programmability. In this paper, we focus on wideband cell-free multi-RIS-empowered Multiple-Input Single-Output (MISO) systems and present a decentralized cooperative active and passive beamforming scheme, aiming to provide an efficient alternative towards the  cooperation overhead of available centralized schemes depending on central processing unit. Considering imperfect channel information availability and realistic frequency selectivity behavior of each RIS's element response, we devise a distributed optimization approach based on consensus updates for the RISs' phase configurations. Our simulation results showcase that the proposed distributed design is superior to centralized schemes that are based on various Lorentzian-type wideband modeling approaches for the RISs.
\end{abstract}


\IEEEpeerreviewmaketitle

\section{Introduction} \label{sec:Intro}
Sixth Generation ($6$G) wireless networks are expected to provide massive connectivity, ultra-high data rates, ultra-low latency, and significant improvements in energy and spectral efficiencies \cite{tataria2021_6g}. In recent years, several novel technologies, such as massive Multiple-Input Multiple-Output (MIMO) and Reconfigurable Intelligent Surfaces (RISs)~\cite{10230036}, have been proposed to fulfill the $6$G vision and treat previous generations' limitations. In fact, according to recent studies \cite{ngo2017cell}, cell-free networks can overcome the limitations of their cellular counterparts, providing high directional gains, bypassing the cell-boundary effect and eliminating inter-cell interference \cite{IBC_distributed2024}. 

However, the paradigm of cell-free networks is generally based on a Central Processing Unit (CPU) requiring high computational resources, which communicates with the deployed base stations (BSs) through dedicated backhaul links~\cite{zhang2021joint}. This centralized approach requires perfect local Channel State Information (CSI) knowledge at the BSs, which should in turn be transmitted to the CPU for beamforming design. Meanwhile, multiple RISs are more beneficial than single RIS-assisted wireless systems \cite{ma2022cooperative}, \cite{al2024performance}, therefore the design of low complexity beamforming algorithms for centralized RISs-assisted cell-free systems becomes a significant challenge. Very recently, an efficient centralized algorithm was proposed in \cite{wang2025efficient}, which takes into account the frequency-selective property of metamaterial elements for Orthogonal Frequency Division Multiplexing (OFDM) transmissions. 

In an effort to resolve computational overhead issues, recent studies proposed alternative decentralized schemes for cell-free systems \cite{huang2020decentralized,xu2023algorithm} and partially distributed beamforming designs \cite{ni2022partially}. Specifically, \cite{huang2020decentralized} proposed an Alternating Direction Method of Multipliers (ADMM) scheme based on the random parallel walk method, and \cite{xu2023algorithm} presented a learning-based deep distributed ADMM framework, with both enforcing complicated consensus constraints for multi-RIS configurations. On the other hand, to avoid the difficulty introduced by such constraints, \cite{ni2022partially} proposed a partially distributed system, according to which the CPU takes responsibility for the optimization of the large-dimensional passive beamformer, and each BS only locally optimizes its reduced dimensionality active beamformer. Nevertheless, all above recent studies focus on perfect CSI knowledge at the BSs and the CPU.

In this paper, we propose a consensus-based distributed beamforming algorithm for wideband cell-free multi-RIS-empowered Multiple-Input Single-Output (MISO) systems that is robust to imperfect CSI availability and is based on successive concave approximation. A novel network-graph-based framework is designed according to which the proposed algorithm can handle inherently the desired consensus for the multi-RIS passive beamforming across the BSs, avoiding the inclusion of additional relevant constraints.
Our simulation results demonstrate that the proposed approach outperforms existing centralized designs, achieving higher rates for both cooperative and non-cooperative beamforming scenarios. 

\textit{Notations:} Boldface lower-case and upper-case letters represent vectors and matrices, respectively. The transpose, Hermitian transpose, conjugate, and the real part of a complex quantity are represented by $(\cdot)^{\rm T}$,  $(\cdot)^{\rm H}$, $(\cdot)^*$, and $\Re\{ \cdot \}$, respectively, while $\mathbb{C}$ is the set of complex numbers, and $\jmath\triangleq\sqrt{-1}$ is the imaginary unit. The symbols $<\!\cdot,\cdot>$ and $\mathbb{E}\{\cdot\}$ denote the inner product and the statistical expectation, respectively, and $\mathbf{x}\sim\mathcal{CN}(\mathbf{a},\mathbf{A})$ indicates a complex Gaussian random vector with mean $\mathbf{a}$ and covariance matrix $\mathbf{A}$. $[\vect{A}]_{i,j}$ represents the $(i,j)$-th element of matrix $\vect{A}$, $\operatorname{diag}\{\mathbf{a}\}$ is defined as the matrix whose diagonal elements are the entries of $\mathbf{a}$, while $\operatorname{vec}_{\rm d}(\vect{A})$ denotes the vector obtained by the diagonal elements of the square matrix $\vect{A}$. $\nabla_{\mathbf{a}}f$ denotes the Euclidean gradient vector of function $f(\cdot)$ along the direction indicated by $\mathbf{a}$.

\section{System Model and Problem Formulation} \label{sec:Sys_Prob_Form}
We consider a wireless communication system comprising $B$ BSs each with $N$ antennas, all wishing to communicate in the downlink direction with $U$ single-antenna User Equipment (UE) via properly placed $R$ identical $M$-element RISs. All BSs send information to all UEs using OFDM in a common set of physical resources, e.g., time and bandwidth. We assume a cell-free structure for the considered system, according to which there is no association among the BSs and the UEs, in contrast to conventional cellular networks. In this way, inter-cell interference becomes irrelevant, and the goal is to remove any noise contribution, improving the achievable Signal-to-Interference-plus-Noise Ratio (SINR). In addition, we assume that the deployed RISs can be shared among the BSs without requiring each RIS to be controlled by a dedicated BS. Thus, there are totally $(R+1)BU$ communicating BS-UE links, among which $BU$ are direct links between BSs and UEs, while the rest correspond to RIS-assisted tunable connections. 

Based on the adopted OFDM scheme, the entire available bandwidth is uniformly distributed across $K$ orthogonal Sub-Carriers (SCs), ensuring that each one occupies an equal spectrum portion. Let $s_{u,k}$ be the transmitted symbol for each $u$-th UE at each $k$-th SC satisfying $\mathbb{E}[\vect{s}_k \vect{s}_k^{\rm H}] = \vect{I}_U$ ($\vect{s}_k \triangleq [s_{1,k},s_{2,k},\dots,s_{U,k}] \in \mathbb{C}^{U\times 1}$), which is digitally processed by the $b$-th BS before transmission with the linear precoding vector $\vect{w}_{b,u,k} \in \mathbb{C}^{N\times 1}$. Then, the total transmitted signal this BS is $\vect{x}_{b,k} \triangleq \sum_{u=1}^U \vect{w}_{b,u,k}s_{u,k}$. We assume that the total transmit power per BS is equal to $P_{b}^{\max}$, and thus the condition $\sum_{u=1}^U \sum_{k=1}^K \norm{\vect{w}_{b,u,k}}^2 \leq P_{b}^{\max}$ must hold.

\subsection{RIS Element Frequency Response} \label{sec:RIS_Response}
In this paper, we adopt the quite general assumption that each $m$-th ($m=1,2,\dots,M$) unit meta-element of each $r$-th ($r=1,2,\dots,R$) RIS can be characterized as an equivalent resonant circuit comprising a resistor ${\rm R_0}$, a tunable capacitor $[\vect{c}_r]_m$ and two inductors $L_1$ and $L_2$ connected in parallel. Then, the frequency response of each $r$-th RIS's $m$-th element can be expressed in the frequency domain as follows~\cite{JSTSP_distributed}:
\begin{equation} \label{eqn:RIS_Scattering}
    [\vect{\phi}_r(f,\vect{c}_r)]_m \triangleq \frac{Z(f,[\vect{c}_r]_m) - \zeta_0}{Z(f,[\vect{c}_r]_m) + \zeta_0},
\end{equation}
where $\zeta_0$ is the characteristic impedance of the propagation medium, while $Z(f,[\vect{c}_r]_m)$ is a complex number describing the characteristic impedance of the equivalent circuit of each $r$-th RIS's element. As a result, letting $\kappa\triangleq 2\pi$, its holds that:
\begin{equation} \label{eqn:impedances}
    Z(f,[\vect{c}_r]_m) = \frac{\jmath\kappa f L_1\left(\jmath\kappa f L_2 + {\rm R_0} + \frac{1}{\jmath\kappa f [\vect{c}_r]_m} \right)}{\jmath\kappa f \left(L_1 + L_2\right) + {\rm R_0} + \frac{1}{\jmath\kappa f [\vect{c}_r]_m}}.
\end{equation}

\subsection{Received Signal Model} \label{sec:Signal_Model}
Let $\vect{h}_{b,u,k} \in \mathbb{C}^{N\times 1}$ denote the direct channel between the $b$-th BS and the $u$-th UE at each $k$-th SC. In addition to direct links, there are also the BS-RIS-UE links, through which the signals transmitted by the $b$-th BS are reflected by the $r$-th RIS before arriving at the $u$-th UE. Let also $\vect{H}_{b,r,k} \in \mathbb{C}^{M\times N}$ and $\vect{g}_{r,u,k}\in\mathbb{C}^{M\times 1}$ represent each BS-RIS channel and each RIS-UE channel, respectively, at each $k$-th SC. Then, the baseband received signal at each $u$-th UE can be mathematically expressed in the frequency domain as $y_{u,k} = \sum_{b=1}^B y_{b,u,k} + n_{u,k}$,
where $y_{b,u,k}$ denotes the received signal through the $b$-th BS, which is given by:
\begin{equation*} 
    y_{b,u,k} = \left(\vect{h}_{b,u,k}^{\rm H} + \sum_{r=1}^R \vect{g}_{r,u,k}^{\rm H} \vect{\Phi}_{r,k}\vect{H}_{b,r,k}\right)\sum_{q=1}^U\vect{w}_{b,q,k}s_{q,k}
\end{equation*}
with $n_{u,k}\sim\mathcal{CN}(0,\sigma_{u,k}^2)$ being the Additive White Gaussian Noise (AWGN) modeling thermal noise at the $u$-th UE, while $\vect{\Phi}_{r,k} \triangleq \operatorname{diag}\{\vect{\phi}_r(f_k,\vect{c}_r)\}$. Using the definition $\vect{f}_{b,u,k}\triangleq \vect{h}_{b,u,k} + \sum_{r=1}^R \vect{H}_{b,r,k}^{\rm H}\vect{\Phi}_{r,k}^{\rm H}\vect{g}_{r,u,k} \in \mathbb{C}^{N\times 1}$, the received signal model can be rewritten more compactly as follows:
\begin{equation*} 
    y_{u,k}=\sum_{b=1}^B\vect{f}_{b,u,k}^{\rm H}\vect{w}_{b,u,k}s_{u,k}+\sum_{\substack{q=1,\\q\neq u}}^U\sum_{b=1}^B \vect{f}_{b,u,k}^{\rm H}\vect{w}_{b,q,k}s_{q,k} + n_{u,k},
\end{equation*}
where the first summation is the desired signal, while the second set of summation terms along with the AWGN term stand for the total interference-plus-noise at the $u$-th UE. 

\subsection{Design Objective and Problem Formulation} \label{sec:Prob_Form}
Based on the received signal model, we define the following vector/matrix of tunable parameters: \textit{i}) the precoding vector $\widetilde{\vect{w}}\triangleq [\tilde{\vect{w}}_1^{\rm T},\dots,\tilde{\vect{w}}_B^{\rm T}]^{\rm T}\in\mathbb{C}^{BUKN\times 1}$ with $\tilde{\vect{w}}_b\triangleq [\vect{w}_{b,1}^{\rm T},\dots,\vect{w}_{b,U}^{\rm T}]^{\rm T}\in\mathbb{C}^{UKN\times 1}$ and $\vect{w}_{b,u}\triangleq [\vect{w}_{b,u,1}^{\rm T},\dots,\vect{w}_{b,u,K}^{\rm T}]^{\rm T} \in \mathbb{C}^{KN\times 1}$; and \textit{ii}) the tunable capacitors $\widetilde{\vect{c}}\triangleq [\vect{c}_1^{\rm T},\vect{c}_2^{\rm T},\dots,\vect{c}_R^{\rm T}]^{\rm T}\in \mathbb{R}^{RM\times 1}$. By treating the Multi-UE Interference (MUI) as an additional source of noise, the instantaneous achievable rate performance per $u$-th UE can be expressed 
as the following function of the system parameters:
\begin{equation*} 
    \mathcal{R}_u(\widetilde{\vect{w}},\widetilde{\vect{c}}) = \sum_{k=1}^K\log_2\left(1\!+\!\frac{\left| \sum\limits_{b=1}^B \vect{f}_{b,u,k}^{\rm H}\vect{w}_{b,u,k}\right|^2}{\sigma_{u,k}^2 + \sum\limits_{\substack{q=1,\\q\neq u}}^U\left|\sum\limits_{b=1}^B \vect{f}_{b,u,k}^{\rm H}\vect{w}_{b,q,k}\right|^2} \right).
\end{equation*}
In this paper, we focus on maximizing the achievable sum rate of the considered wideband cell-free multi-RIS-empowered MISO system under imperfect CSI availability. In mathematical terms, we formulate the following optimization problem:
\begin{align*}
	\mathcal{OP}: \,\max_{\widetilde{\vect{w}},\widetilde{\vect{c}}} \, & \quad \mathbb{E}_{\vect{h},\vect{g},\vect{H}} \left[\sum_{u=1}^U \mathcal{R}_u\left(\widetilde{\vect{w}},\widetilde{\vect{c}}\right)\right] \\
	\text{s.t.} & \quad \sum_{u=1}^{U}\sum_{k=1}^{K} \norm{\vect{w}_{b,u,k}}^2 \leq P_b^{\max} \;\forall b \in \mathscr{B}\\
    & \quad C_{\min} \leq [\vect{c}_r]_{m} \leq C_{\max}  \;\forall m\in\mathscr{M},
\end{align*}
where the expectation in the objective function is taken with respect to estimation errors for the channel gains encapsulated into $\vect{f}_{b,u,k}$ for all indices $b$, $u$, $k$, as well as set $\mathscr{B}\triangleq\{1,2,\ldots,B\}$, $\mathscr{M}\triangleq\{1,2,\ldots,M\}$, and $\mathscr{R}\triangleq\{1,2,\ldots,R\}$. Finally, parameters $C_{\min}$ and $C_{\max}$ denote the permissible minimum and maximum values for the RIS tunable capacitors. 

\section{Distributed Sum-Rate Maximization} \label{sec:Distr_OP_Solution}
In this paper, our aim is to solve $\mathcal{OP}$ in a decentralized manner, treating BSs as agents whose objective is to maximize the system's sum-rate performance cooperatively, with minimum exchange messages. However, the objective function cannot be decomposed with respect to each BS, a feature that usually arises in the cellular case \cite{JSTSP_distributed,IBC_distributed2024}. This happens because the objective for the cell-free case is user-centric, a fact that produces inherent difficulties to treat $\mathcal{OP}$ in a decentralized way. In the sequel, we present a distributed beamforming framework tailored to the considered shared utilization of the multiple RISs among the system's BSs. In particular, the goal is to design the RIS variables $\widetilde{\vect{c}}$ such that all BSs agree on them, while the linear precoders included in $\tilde{\vect{w}}_{b}$ will be decided locally at each BS side.

Based on the above observations and capitalizing on the stochastic decomposition method detailed in \cite{yang2016parallel}, $\mathcal{OP}$ can be solved in a distributed manner, as follows. Each $b$-th BS solves iteratively $\mathcal{OP}_{\mathcal{D},b}: \max_{\vect{x}_b \in \mathcal{X}} \widehat{\mathcal{R}}\left(\vect{x}_b;\vect{x}_{\bar b}^t,\vect{\eta}_b^t\right)$,
where $\vect{x}_b \triangleq \left[\tilde{\vect{w}}_b^{\rm T},\widetilde{\vect{c}}_b^{\rm T}\right]^{\rm T}$ includes the design parameters decided by each $b$-th BS with $\mathcal{X}$ denoting their feasible set presented in $\mathcal{OP}$'s constraints, while, for each $t$-th algorithmic iteration, $\vect{x}_{\bar{b}}^t$ collects all other BSs' design parameters and $\vect{\eta}_b^t$ represents the locally generated noisy channel samples. The objective function of each $\mathcal{OP}_{\mathcal{D},b}$ can be further expressed as:
\begin{align}\label{eqn:rate_stoch_surrogate}
    &\widehat{\mathcal{R}}(\vect{x}_b;\vect{x}_{\bar{b}}^t,\vect{\eta}_b^t) \triangleq \,\rho^t\widetilde{\mathcal{R}}_u(\vect{x}_b;\vect{x}_{\bar{b}}^t,\vect{\eta}_b^t) + \rho^t<\vect{\pi}_{\vect{x}_b}^t,\vect{x}_b - \vect{x}_b^t>\nonumber \\
    &\qquad\quad+(1- \rho^t)<\vect{d}_{\vect{x}_b}^{t},\vect{x}_b - \vect{x}_b^t> 
    - \frac{\tau}{2}\norm{\vect{x}_b - \vect{x}_b^t}^2,
\end{align}
where $\widetilde{\mathcal{R}}_u(\vect{x}_b;\vect{x}_{\bar{b}}^t,\vect{\eta}_b^t)\triangleq\mathcal{R}_u(\vect{x}_b;\vect{x}_{\bar{b}}^t,\vect{\eta}_b^t)\,\,\,+<\vect{\gamma}_{\widetilde{\vect{c}}_b}^t,\widetilde{\vect{c}}_b - \widetilde{\vect{c}}_b^t>$, with $\vect{\gamma}_{\widetilde{\vect{c}}_b}^t \triangleq \nabla_{\widetilde{\vect{c}}_b}\mathcal{R}_u(\widetilde{\vect{c}}_b^t;\vect{x}_{\bar{b}}^t,\vect{\eta}_b^t)$ having role to decouple $\tilde{\vect{w}}_b$ and $\widetilde{\vect{c}}_b$ in the objective function, $\vect{\pi}_{\vect{x}_b}^t \triangleq \sum_{q=1,q\neq u}^U \nabla_{\vect{x}_b}\mathcal{R}_q(\vect{x}_b;\vect{x}_{\bar{b}}^t,\vect{\eta}_b^t)$ is the pricing vector, $\tau$ is a positive constant, and the last term in \eqref{eqn:rate_stoch_surrogate} denotes the proximal regularization to ensure strong concavity. On the other hand, $\vect{d}_{\vect{x}_b}^t$ serves as an online estimate of the gradient (also called accumulation vector, see, e.g., \cite{yang2016parallel}) of the objective function for $\mathcal{OP}$ at the realization $\vect{\eta}_b^t$, which is updated recursively as:
\begin{equation} \label{eqn:accum_vec_def}
    \vect{d}_{\vect{x}_b}^t = (1 - \rho^t)\vect{d}_{\vect{x}_b}^{t-1} + \rho^t\left( \vect{\pi}_{\vect{x}_b}^t + \nabla_{\vect{x}_b}\mathcal{R}_u(\vect{x}_b^t;\vect{x}^t,\vect{\eta}_b^t) \right),
\end{equation}
and $\rho^t \in (0,1]$ is a suitably chosen step-size sequence (with $\rho^0 = 1$). In the next subsections, we solve $\mathcal{OP}_{\mathcal{D},b}$ locally for each of the two sets of design parameters, i.e., $\tilde{\vect{w}}_b$ and $\widetilde{\vect{c}}_b$.

\subsection{Precoding Optimization at Each $b$-th BS} \label{sec:Precoder_Design}
Writing $\mathcal{OP}_{\mathcal{D},b}$ with respect to the linear precoder of the $b$-th BS reduces to the following formulation:
\begin{align*}
    \mathcal{OP}_{\vect{w}_b}: \,\max_{\vect{w}_{b,u,k}} \, & \quad \rho^t\widetilde{\mathcal{R}}_u + \rho^t<\vect{\pi}_{\vect{w}_{b,u,k}}^t,\vect{w}_{b,u,k} - \vect{w}_{b,u,k}^t> \\
    &+(1-\rho^t)<\vect{d}_{\vect{w}_{b,u,k}}^{t},\vect{w}_{b,u,k} - \vect{w}_{b,u,k}^t> \\
    &- \frac{\tau}{2}\norm{\vect{w}_{b,u,k} - \vect{w}_{b,u,k}^t}^2 \\
	\text{s.t.} & \quad \sum_{u=1}^{U}\sum_{k=1}^{K} \norm{\vect{w}_{b,u,k}}^2 \leq P_b^{\max}.
\end{align*}
To proceed with the solution of $\mathcal{OP}_{\vect{w}_b}$, we first treat the non-concave term $\widetilde{\mathcal{R}}_u$. To this end, we use \cite[Lemma 1]{JSTSP_distributed} and references therein, according to which $\widetilde{\mathcal{R}}_u$ can be lower-bounded by $\sum_{k=1}^K \breve{\mathcal{R}}_{u,k}$, where $\breve{\mathcal{R}}_{u,k}$ can be derived as:
\begin{equation} \label{eqn:precod_surrogate}
\begin{aligned}
    \breve{\mathcal{R}}_{u,k} \triangleq &-c_{u,k}^t \left|\sum_{b=1}^B \vect{f}_{b,u,k}^{\rm H} \vect{w}_{b,u,k}\right|^2 \\
    &+ 2\Re\left\{(e_{u,k}^t)^*\left(\sum_{b=1}^B \vect{f}_{b,u,k}^{\rm H} \vect{w}_{b,u,k}\right)\right\},
\end{aligned}    
\end{equation}
where we have used the following definitions:
\begin{align} \label{eqn:surrog_defs}
    c_{u,k}^t &\triangleq \frac{1}{\ln(2)}\frac{|\beta_{u,k}^t|^2}{\alpha_{u,k}^t(\alpha_{u,k}^t - |\beta_{u,k}^t|^2)}, \\
    e_{u,k}^t &\triangleq \frac{1}{\ln(2)} \frac{(\alpha_{u,k}^t)^{-1} \beta_{u,k}^t}{1 - \alpha_{u,k}^{-1}|\beta_{u,k}^t|^2}, \\
    \alpha_{u,k}^t &\triangleq \sigma_{u,k}^2 + \sum_{q=1,q\neq u}^U\left|\sum_{b=1}^B\vect{f}_{b,u,k}^{\rm H}\vect{w}_{b,q,k}^t \right|^2 + \left|\beta_{u,k}^t\right|^2, \\
    \beta_{u,k}^t &\triangleq \sum_{b=1}^B \vect{f}_{b,u,k}^{\rm H}\vect{w}_{b,u,k}^t.    
\end{align}
By letting $r_{u,k}^t \triangleq \sum_{b'=1,b'\neq b}^B\vect{f}_{b',u,k}^{\rm H}\vect{w}_{b',u,k}^t$, \eqref{eqn:precod_surrogate}'s surrogate function can be manipulated as a function of $\vect{w}_{b,u,k}$:
\begin{equation}
\begin{aligned}
    \breve{\mathcal{R}}_{u,k} = &-c_{u,k}^t\vect{w}_{b,u,k}^{\rm H}\vect{f}_{b,u,k}\vect{f}_{b,u,k}^{\rm H}\vect{w}_{b,u,k} \\
    &+ 2\Re\left\{(e_{u,k}^t - c_{u,k}^t r_{u,k}^t)^*\vect{f}_{b,u,k}^{\rm H}\vect{w}_{b,u,k}\right\},
\end{aligned}    
\end{equation}
which is a concave function with respect to $\vect{w}_{b,u,k}$ since it is the sum of a concave quadratic plus a linear term.

Next, we derive the necessary expressions for the pricing vectors $\vect{\pi}_{\vect{w}_{b,u,k}}^t$ and accumulation vectors $\vect{d}_{\vect{w}_{b,u,k}}^t$. Let, first, 
\begin{align}
    \operatorname{snr}_{q,k}^t &\triangleq \frac{\abs*{\sum\limits_{b=1}^B\vect{f}_{b,q,k}^{\rm H}\vect{w}_{b,q,k}^t}^2}{\operatorname{MUI}_{q,k}^t}, \\
    \operatorname{MUI}_{q,k}^t &\triangleq \sigma_{q,k}^2 + \sum\limits_{\substack{u=1,\\u\neq q}}^U \abs*{\sum\limits_{b=1}^B \vect{f}_{b,q,k}^{\rm H}\vect{w}_{b,u,k}^t}^2.
\end{align}
Then, it can be shown that the pricing vector is given by:
\begin{equation} \label{eqn:pricing_w}
    \vect{\pi}_{\vect{w}_{b,u,k}}^t = \frac{-1}{\ln(2)}\sum\limits_{\substack{q=1,\\q\neq u}}^U\frac{\operatorname{snr}_{q,k}^t\left(\vect{f}_{b,q,k}^{\rm H}\vect{w}_{b,u,k}^t + r_{q,u,k}^t \right)}{(1 + \operatorname{snr}_{q,k}^t)\operatorname{MUI}_{q,k}^t}\vect{f}_{b,q,k},
\end{equation}
while, for the accumulation vector, it holds that:
\begin{equation} \label{eqn:accum_w}
\begin{aligned}
    \vect{d}_{\vect{w}_{b,u,k}}^t = &\,(1 - \rho^t)\vect{d}_{\vect{w}_{b,u,k}}^{t-1} + \rho^t\Big( \vect{\pi}_{\vect{w}_{b,u,k}}^t \\
    &+ \frac{1}{\ln(2)}\frac{\vect{f}_{b,u,k}^{\rm H}\vect{w}_{b,u,k}^t + r_{u,k}^t}{(1 + \operatorname{snr}_{u,k}^t)\operatorname{MUI}_{u,k}^t}\vect{f}_{b,u,k}\Big),
\end{aligned}
\end{equation}
where $r_{q,u,k}^t \triangleq \sum_{b'=1,b'\neq b}^B\vect{f}_{b',q,k}^{\rm H}\vect{w}_{b',u,k}^t$. Based on the above derivations, it can be deduced that $\mathcal{OP}_{\vect{w}_b}$ is a concave optimization problem, which can be solved by considering its Lagrangian, which is equal to the following expression:
\begin{align}
    \mathcal{L}(\vect{w}_{b,u,k},\lambda_b) \triangleq &-\vect{w}_{b,u,k}^{\rm H}\vect{Q}_{b,u,k}^t\vect{w}_{b,u,k} + \Re\left\{(\vect{q}_{b,u,k}^t)^{\rm H}\vect{w}_{b,u,k} \right\}\nonumber \\
    &- \lambda_b\left(\sum_{u=1}^U \sum_{k=1}^K \norm{\vect{w}_{b,u,k}}^2 - P_{b}^{\max}\right),
\end{align}
where $\lambda_b \geq 0$ is the Lagrangian multiplier associated with the transmit power constraint and 
\begin{align}
   \vect{Q}_{b,u,k}^t &\triangleq \rho^t c_{u,k}^t \vect{f}_{b,u,k}\vect{f}_{b,u,k}^{\rm H} + \frac{\tau}{2}\vect{I}_N, \\
        \vect{q}_{b,u,k}^t &\triangleq 2(e_{u,k}^t - c_{u,k}^t r_{u,k}^t)\vect{f}_{b,u,k} + \rho^t\vect{\pi}_{\vect{w}_{b,u,k}}^t\nonumber \\
        &\hspace{0.33cm}+ (1-\rho^t)\vect{d}_{\vect{w}_{b,u,k}}^t + \tau\vect{w}_{b,u,k}^t.   
\end{align}
Then, it can be trivially shown that the solution for $\vect{w}_{b,u,k}$ is:
\begin{equation} \label{eqn:optimal_w}
    \vect{w}_{b,u,k}^{\rm opt}(\lambda_b) = \frac{1}{2}\left( \vect{Q}_{b,u,k}^t + \lambda_b\vect{I}_N \right)^{-1} \vect{q}_{b,u,k}^t.
\end{equation}


\subsection{RIS Phase Configuration Optimization at Each $b$-th BS} \label{sec:RIS_Design}
The reduced optimization problem $\mathcal{OP}_{\mathcal{D},b}$ with respect to the capacitors' vectors $\widetilde{\vect{c}}_b$, and eventually the RIS passive beamforming matrices $\vect{\Phi}_{r,k}$ $\forall r\in\mathscr{R}$, that needs to be solved by each BS $b$, is expressed as (letting $\mathscr{N} \triangleq \{1,2,\dots,RM\}$):
\begin{align*}
	\mathcal{OP}_{\widetilde{\vect{c}}_b}: \,\max_{\widetilde{\vect{c}}_b} \, & \quad f(\widetilde{\vect{c}}_b) \triangleq \rho^t<\vect{\gamma}_{\widetilde{\vect{c}}_b}^t + \vect{\pi}_{\widetilde{\vect{c}}_b}^t,\widetilde{\vect{c}}_b-\widetilde{\vect{c}}_b^t> \\
    &+ (1-\rho^t)<\vect{d}_{\widetilde{\vect{c}}_b}^{t},\widetilde{\vect{c}}_b-\widetilde{\vect{c}}_b^t> - \frac{\tau}{2}\norm{\widetilde{\vect{c}}_b - \widetilde{\vect{c}}_b}^2 \\
	\text{s.t.} & \quad C_{\min} \leq [\widetilde{\vect{c}}_b]_n \leq C_{\max} \, \forall n\in\mathscr{N}.
\end{align*} 
To achieve consensus among the deployed BSs, we capitalize on dynamic first-order average methods \cite{xin2020general} which allow tracking the average gradient based on local information exchange. Specifically, as proposed in \cite{lorenzo2016next}, let the average gradient be denoted as $\overline{\nabla\mathcal{R}}(\widetilde{\vect{c}}_b^t;\vect{\eta}^t) \triangleq \frac{1}{B} \nabla_{\widetilde{\vect{c}}_b}\overline{\mathcal{R}}(\widetilde{\vect{c}}_b^t;\vect{\eta}^t)$, where $\overline{\mathcal{R}}(\widetilde{\vect{c}}_b^t;\vect{\eta}^t) \triangleq \sum_{u=1}^U\mathcal{R}_u(\widetilde{\vect{c}}_b^t;\vect{\eta}^t)$, and let also $\vect{q}_{\widetilde{\vect{c}}_b}^t$ be a local auxiliary variable for each $b$-th BS, whose role is to asymptotically track $\overline{\nabla\mathcal{R}}(\widetilde{\vect{c}}_b^t;\vect{\eta}^t)$. Then, according to the dynamic average consensus method in~\cite{zhu2010discrete}, gradient tracking can be enabled based on the following update rule:
\begin{equation*} 
    \vect{q}_{\widetilde{\vect{c}}_b}^{t+1} = \sum_{i\in\mathcal{N}_b^t} [\vect{V}^t]_{b,i} \vect{q}_{\widetilde{\vect{c}}_i}^t + \nabla_{\widetilde{\vect{c}}_b}\overline{\mathcal{R}}(\widetilde{\vect{c}}_b^{t+1};\vect{\eta}^{t+1}) - \nabla_{\widetilde{\vect{c}}_b}\overline{\mathcal{R}}(\widetilde{\vect{c}}_b^{t};\vect{\eta}^{t})
\end{equation*}
with $\vect{q}_{\widetilde{\vect{c}}_b}^{0} = \nabla_{\widetilde{\vect{c}}_b}\overline{\mathcal{R}}(\widetilde{\vect{c}}_b^{0};\vect{\eta}^{0})$. In addition, the entries of the matrix $\vect{V}^t \in \mathbb{R}^{B\times B}$ represent the (possibly time-varying) weights of the connected network graph $\mathcal{G}=(\mathcal{V},\mathcal{E})$, where $\mathcal{V}=\mathscr{B}$ and $\mathcal{E}$ is the set of edges of the network topology considered, assuming that $\vect{V}\vect{1}=\vect{1}$ and $\vect{1}^{\rm T}\vect{V}^{\rm T}=\vect{1}^{\rm T}$, i.e., double stochasticity must hold for $\vect{V}$. Also $\mathcal{N}_b^t$ denotes the neighbors of BS $b$. Hence, given $\vect{q}_{\widetilde{\vect{c}}_b}^{t+1}$, BS $b$ can locally compute its pricing vector $\vect{\pi}_{\widetilde{\vect{c}}_b}^t$ based on the expression $\vect{\pi}_{\widetilde{\vect{c}}_b}^{t+1} = B\vect{q}_{\widetilde{\vect{c}}_b}^{t+1} - \nabla_{\widetilde{\vect{c}}_b}\overline{\mathcal{R}}(\widetilde{\vect{c}}_b^{t+1};\vect{\eta}^{t+1})$, while $\vect{d}_{\widetilde{\vect{c}}_b}^{t+1} = (1-\rho^t)\vect{d}_{\widetilde{\vect{c}}_b}^{t} + \rho^t B\vect{q}_{\widetilde{\vect{c}}_b}^{t+1}$. 

Therefore, to proceed with the solution of $\mathcal{OP}_{\widetilde{\vect{c}}_b}$, it suffices to derive the gradient of $\mathcal{R}_u$ with respect to $\widetilde{\vect{c}}_b$, which is necessary for the calculation of $\vect{\gamma}_{\widetilde{\vect{c}}_b}^t$, $\vect{\pi}_{\widetilde{\vect{c}}_b}^t$, and $\vect{d}_{\widetilde{\vect{c}}_b}^t$. To this end, we define the matrices $\vect{A}_{b,q,k} \triangleq \vect{H}_{b,k}\vect{w}_{b,q,k}\vect{w}_{b,q,k}^{\rm H}\vect{H}_{b,k}^{\rm H}$, $\tilde{\vect{A}}_{b,k}\triangleq\sum_{q=1,q\neq u}^U\vect{A}_{b,q,k}$, $\vect{B}_{b,u,q,k} \triangleq \vect{H}_{b,k}\vect{w}_{b,q,k}\vect{w}_{b,q,k}^{\rm H}\vect{h}_{b,u,k}\vect{g}_{u,k}^{\rm H} + r_{u,q,k}^*\vect{H}_{b,k}\vect{w}_{b,q,k}\vect{g}_{u,k}^{\rm H}$, $\tilde{\vect{B}}_{b,u,k}\triangleq\sum_{q=1,q\neq u}^U\vect{B}_{b,u,q,k}$, and $\vect{G}_{u,k} \triangleq \vect{g}_{u,k}\vect{g}_{u,k}^H$, where $\vect{H}_{b,k}\triangleq [\vect{H}_{b,1,k}^{\rm T},\dots,\vect{H}_{b,R,k}^{\rm T}]^{\rm T}\in \mathbb{C}^{RM\times N}$ and $\vect{g}_{u,k}\triangleq [\vect{g}_{1,u,k}^{\rm T},\dots,\vect{g}_{R,u,k}^{\rm T}]^{\rm T} \in \mathbb{C}^{RM\times 1}$. Let also $f_{1,u,k}\triangleq \left| \sum_{b=1}^B\vect{f}_{b,u,k}^{\rm H}\vect{w}_{b,u,k}^t \right|^2$ and $f_{2,u,k}\triangleq \operatorname{MUI}_{u,k}$. Then, based on \cite[Th. 1]{JSTSP_distributed}, it can be shown that:
\begin{equation} \label{eqn:overall_gradient_cap}
\begin{aligned}
    \nabla_{\widetilde{\vect{c}}_b}\mathcal{R}_u = &\frac{1}{\ln(2)}\sum_{k=1}^K \Biggl(\frac{1}{(1 + \frac{f_{1,u,k}}{f_{2,u,k}})f_{2,u,k}^2}\\
    &\times\left( f_{2,u,k}\nabla_{\widetilde{\vect{c}}_b}f_{1,u,k} - f_{1,u,k}\nabla_{\widetilde{\vect{c}}_b}f_{2,u,k} \right)\Biggr),
\end{aligned}
\end{equation}
where the gradient vectors in this expression are given by:
\begin{align}
    \nabla_{\widetilde{\vect{c}}_b}f_{1,u,k} &= 
        \vect{D}_{\widetilde{\vect{c}}_b}^*\operatorname{vec}_{\rm d}(\vect{G}_{u,k}\vect{\Phi}_k\vect{A}_{b,u,k} + \vect{B}_{b,u,q,k}^*) \nonumber \\
        &\phantom{=}+ \vect{D}_{\widetilde{\vect{c}}_b}\operatorname{vec}_{\rm d}(\vect{G}_{u,k}^{\rm T}\vect{\Phi}_k^*\vect{A}_{b,u,k}^{\rm T} + \vect{B}_{b,u,q,k}^{\rm T}), \label{eqn:grad_f1} \\
    \nabla_{\widetilde{\vect{c}}_b}f_{2,u,k} &= 
        \vect{D}_{\widetilde{\vect{c}}_b}^*\operatorname{vec}_{\rm d}(\vect{G}_{u,k}\vect{\Phi}_k\tilde{\vect{A}}_{b,k} + \tilde{\vect{B}}_{b,u,k}^*) \nonumber \\
        &\phantom{=}+ \vect{D}_{\widetilde{\vect{c}}_b}\operatorname{vec}_{\rm d}(\vect{G}_{u,k}^{\rm T}\vect{\Phi}_k^*\tilde{\vect{A}}_{b,k}^{\rm T} + \tilde{\vect{B}}_{b,u,k}^{\rm T}), \label{eqn:grad_f2}
\end{align}
while $\vect{D}_{\widetilde{\vect{c}}_b}\in\mathbb{C}^{RM\times RM}$ is a diagonal matrix with 
\begin{equation} \label{eqn:grad_capacitors}
    [\vect{D}_{\widetilde{\vect{c}}_b}]_{n,n} \triangleq 2\zeta_0\frac{\frac{\partial\mathcal{N}_n}{\partial([\widetilde{\vect{c}}_b]_n)}\mathcal{D}_n - \mathcal{N}_n\frac{\partial\mathcal{D}_n}{\partial([\widetilde{\vect{c}}_b]_n)}}{(\mathcal{N}_n + \zeta_0\mathcal{D}_n)^2}\,\forall n\in\mathscr{N},
\end{equation}
$\mathcal{N}_n$ and $\mathcal{D}_n$ are equal to the numerator and denominator of \eqref{eqn:impedances}, respectively, evaluated at the $k$-th SC, as well as $\frac{\partial\mathcal{N}_n}{\partial([\widetilde{\vect{c}}_b]_n)}=\frac{-L_1}{([\widetilde{\vect{c}}_b]_n)^2}$ and $\frac{\partial\mathcal{D}_n}{\partial([\widetilde{\vect{c}}_b]_n)}=\frac{\jmath}{\kappa f_k([\widetilde{\vect{c}}_b]_n)^2}$. Finally, $\mathcal{OP}_{\widetilde{\vect{c}}_b}$ can be solved in closed form following \cite[Cor.~1]{JSTSP_distributed}. 

\subsection{Overall Distributed Solution} \label{sec:Proposed_Design}
In the previous two subsections, the best-response mapping $\hat{\vect{x}}_b$ was calculated for the BS linear precoders and the RIS phase configurations, by solving $\mathcal{OP}_{\vect{w}_b}$ and $\mathcal{OP}_{\widetilde{\vect{c}}_b}$, respectively. Next, we use the rule $\vect{x}_b^{t+1} = (1-\alpha^t)\vect{x}_b^t +\alpha^t\hat{\vect{x}}_b$ to update both parameters, while, for $\widetilde{\vect{c}}_b$, one more step is necessary to achieve consensus, which is given by $\widetilde{\vect{c}}_b^{t+1} = \sum_{i\in\mathcal{N}_b^t} [\vect{V}]_{b,i}^t \breve{\vect{c}}_i^{t+1}$. The detailed algorithmic steps for solving $\mathcal{OP}$ will be described in the extended version of this work.

\section{Numerical Results and Discussion} \label{sec:Num_Results}
In this section, we investigate the performance of the algorithmic solution proposed to tackle $\mathcal{OP}$. 
For the experimental setup, we have considered that all nodes were positioned on the $3$-Dimensional ($3$D) Cartesian coordinate system: $B=4$ BSs, with each $b$-th BS located in $(50(b-1)\,m,0\,m,5\,m)$ and equipped with $N=2$ antennas while serving $U=4$ UEs. The UEs were divided and randomly placed into two separate circular groups, each with the same radius equal to $2\,m$. In particular, the center of the first cluster was placed on the $x$-axis at $67.5\,m$ and the second at $82.5\,m$, sharing the same $y$-coordinate at $57.5\,m$, and the $z$-coordinate for each UE was set equal to $1.5\,m$. The cell-free system also included $R=2$ RISs each equipped with $M=144$ unit elements, whose coordinates were fixed to $(65,60,6)\,m$ for $\rm{RIS_1}$ and $(85,60,6)$ for $\rm{RIS_2}$. All wireless links were modeled as wideband Rayleigh fading channels with distance-dependent pathloss, which was modeled as $\operatorname{PL}_{i,j} = \operatorname{PL}_0(d_{i,j}/d_0)^{-\alpha_{i,j}}$ with $\operatorname{PL}_0=-30\,{\rm dB}$, $d_{i,j}$ being the distance between any two nodes $i,j$ ($(i,j)\in\{\rm{BS},\rm{UE},\rm{RIS}\}$), and $d_0 = 1\,m$. For pathloss exponents, we have set $\alpha_{\rm{BS,UE}} = 3.8$, $\alpha_{\rm{BS,RIS}} = 2.4$, and $\alpha_{\rm{RIS,UE}} = 2.2$, while to produce imperfect channel samples (i.e., $\vect{\eta}^t)$, we have used the model described in \cite[eq. (53)]{zhang2021joint}, according to which $\hat{g} = g + e$ with $g$ being the actually estimated channel gain and $e\sim\mathcal{CN}(0,\sigma_e^2)$, $\sigma_e^2\triangleq\delta\abs{g}^2$ and $\delta=0.2$. In addition, in the performance results that follow, we have set equal transmit power $P_b^{\max}=P_{\max}$ $\forall b\in\mathscr{B}$ and noise variance $\sigma_{u,k}^2 = \sigma^2=-90$ dBm at all UEs. Also the carrier frequency was set as $f_c=3.5$ GHz, the bandwidth $\rm{BW}=100$ MHz, and $K=16$ SCs. The circuital elements for each RIS were set as follows: $L_1 = 1.7143$ nH, $L_2 = 0.48$ nH, ${\rm R_0} = 1\,{\rm \Omega}$, $\zeta_0 = 50\,\rm{\Omega}$, $C_{\min} = 0.01$ pF, and $C_{\max} = 3$ pF. We have set the algorithmic parameters: $\tau=10^{-2}$ and $\epsilon = 10^{-3}$ (convergence threshold), while, for the step sizes $\rho^t$ and $\alpha^t$, we have used the updating rules $\rho^t = \frac{1}{(t+2)^{0.99}}$, and $\alpha^t=\frac{1}{t+2}$. In the results that follow, we have used $100$ independent channel realizations. We also considered that all BSs communicate with each other and built $\vect{V}$ based on the Metropolis-Hastings weights \cite{lorenzo2016next}.
\begin{figure}[!t]
	\includegraphics[width=3.20in]{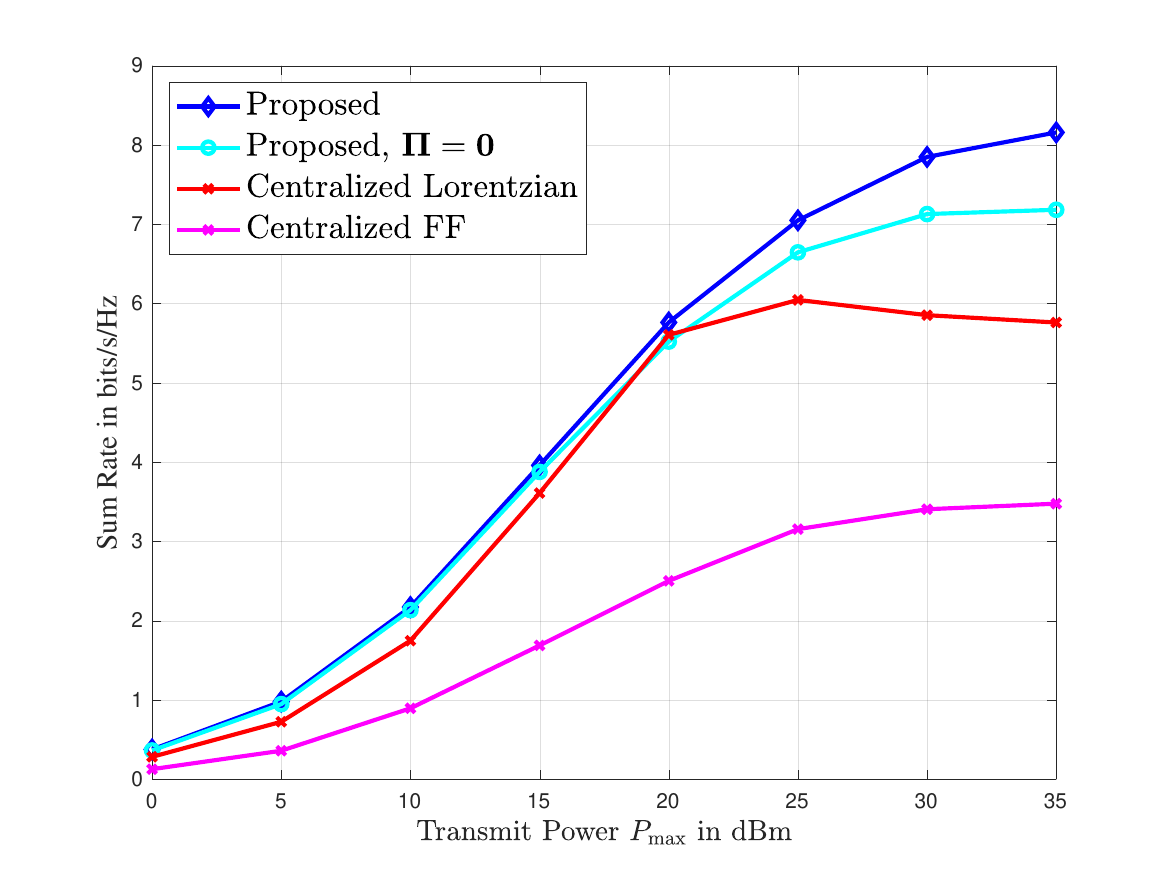}
	\caption{\small{Achievable sum-rate performance for the considered wideband cell-free MISO system with $B=4$ BSs, each with $N=2$ antennas, and $R=2$ RISs with $M=144$ unit elements each.\vspace{-0.3cm}}}
	\label{fig:Rates_vs_P}
\end{figure}
For comparison purposes, we have implemented the centralized algorithms described in \cite{zhang2021joint} (termed as ``Centralized FF'') and \cite{wang2025efficient} (named ``Centralized Lorentzian''), which were designed considering perfect CSI knowledge, but, for a fair comparison with the proposed framework, they were evaluated over the previously described erroneous channels. Note that \cite{zhang2021joint}'s design does not consider frequency selective RISs, hence, we used its output solution denoted by $\vect{\phi}_r^{\rm centr}$ and solved the equation \eqref{eqn:RIS_Scattering} with respect to $\vect{c}_r$ numerically, letting $f=f_c$ and the left-hand-side equal to $\vect{\phi}_r^{\rm centr}$ $\forall r$. Finally, we also examined the case of no cooperation ($\vect{\Pi}=\vect{0}$) among the BSs (refer to \cite{JSTSP_distributed} for details).

In Fig.~\ref{fig:Rates_vs_P}, we illustrate the performance of all aforedescribed schemes versus $P_{\max}$. It can be observed that all depicted curves follow a non-decreasing trend as $P_{\max}$ increases. More importantly, the proposed design is shown to outperform the ``Centralized Lorentzian'' scheme, achieving higher performance, especially when $P_{\max}>20$ dBm, which can be explained by the fact that the latter design utilizes a different model for the frequency selective behavior of each metamaterial element. In parallel, the advantages of cooperative beamforming are also pronounced for this range of $P_{\max}$ values, between the two proposed decentralized schemes.

\section{Conclusion} \label{sec:conclusion}
In this paper, we studied cell-free multi-RIS-aided MISO OFDM system under imperfect CSI knowledge, and focused on the sum-rate maximization problem which was tackled in a decentralized manner, in contrast to the widely considered centralized approaches requiring a CPU. It was showcased that the proposed consensus-based distributed design for the BS precoders and the RIS configurations outperforms all benchmark centralized schemes that consider various Lorentzian-type frequency selectivity models for the deployed RISs.

\bibliographystyle{IEEEtran}
\bibliography{References}

\end{document}